\documentclass[pra,reprint,superscriptaddress,floatfix]{revtex4-2}

 \usepackage{comment}
 \usepackage{amsmath}
 \usepackage{amsfonts}
 \usepackage{float}
 \usepackage{graphicx}
  \usepackage{bm}
 \usepackage{comment}
 \usepackage{booktabs}
 \usepackage{newtxtext,newtxmath}
\usepackage{dcolumn}

 \usepackage{hyperref}
 \hypersetup{
    colorlinks=true,       
    linkcolor=blue,          
    citecolor=blue,        
    filecolor=magenta,      
    urlcolor=blue           
}

\renewcommand{\epsilon}{\varepsilon}

\allowdisplaybreaks

\newcolumntype{d}[1]{D{.}{.}{#1}}

\let\originalleft\left
\let\originalright\right
\renewcommand{\left}{\mathopen{}\mathclose\bgroup\originalleft}
\renewcommand{\right}{\aftergroup\egroup\originalright}

\usepackage{adjustbox}

\begin{document}

\frenchspacing

\title{Gaussian-basis many-body theory calculations of positron binding to negative ions and atoms}
\author{J. Hofierka}
\email{jhofierka01@qub.ac.uk}
\affiliation{
School of Mathematics and Physics, Queen's University Belfast, University Road, Belfast BT7 1NN, United Kingdom}
\author{B. Cunningham}
\affiliation{
School of Mathematics and Physics, Queen's University Belfast, University Road, Belfast BT7 1NN, United Kingdom}
\author{C. M. Rawlins}
\affiliation{
School of Mathematics and Physics, Queen's University Belfast, University Road, Belfast BT7 1NN, United Kingdom}
\author{C.~H.~Patterson}
\affiliation{
School of Physics, Trinity College Dublin, Dublin 2, Ireland
}
\author{D.~G. Green}
\email{d.green@qub.ac.uk}
\affiliation{
School of Mathematics and Physics, Queen's University Belfast, University Road, Belfast BT7 1NN, United Kingdom}

\date{\today}

\begin{abstract}
\noindent Positron binding energies in the negative ions H$^-$, F$^-$, Cl$^-$ and Br$^-$, and the closed-shell atoms Be, Mg, Zn and Ca, are calculated via a many-body theory approach developed by the authors [J.~Hofierka \emph{et al.} Nature~{\bf 608}, 688-693 (2022)].
Specifically, the Dyson equation is solved using a Gaussian basis, with the positron self energy constructed from three infinite classes of diagrams 
that account for the strong positron-atom correlations that characterise the system including the positron-induced polarization of the electron cloud, screening of the electron-positron Coulomb interaction, virtual-positronium formation and electron-hole and positron-hole interactions. 
For the negative ions, binding occurs at the static level of theory,  and the correlations are found to enhance the binding energies by $\sim$25--50\%, yielding results in good agreement with ($\lesssim$5\% larger than) calculations from a number of distinct methods.
For the atoms, for which binding is enabled exclusively by correlations, most notably virtual-Ps formation, 
 the binding energies are found to be of similar order to (but $\sim$10--30\% larger than) relativistic coupled-cluster calculations of [C. Harabati, V.~A.~Dzuba and V.~V. Flambaum, Phys.~Rev.~A {\bf 89}, 022517 (2014)], both of which are systematically larger than stochastic variational calculations of [M.~Bromley and J.~Mitroy, Phys.~Rev.~A {\bf 73} (2005); J.~Mitroy, J.~At.~Mol.~Sci.~{\bf 1}, 275 (2010)].
\end{abstract}

\maketitle

\section{Introduction}
The ability of positrons to bind to negative ions has been known for many decades, see, e.g., \cite{Ore:1951,Simons1953,schradermoxom2001}.
Accurate variational calculations have been performed  for the hydrogen anion (the simplest atomic system capable of binding the positron beyond positronium, the electron-positron bound state) \cite{Frolov1997} while calculations for heavier, many-electron ions are less accurate \cite{schradermoxom2001}.  
Most of the attention has been focussed on positronium halides with methods including model potential quantum Monte Carlo \cite{Schrader1992,Schrader1993},  multi-reference configuration interaction \cite{Saito03,Saito05},  atomic many-body theory \cite{Ludlow10}, the APMO method \cite{Romero:2014}, and diffusion Monte Carlo \cite{bressanini98,Ito2020,Charry2022}.

The ability of certain neutral atoms to support positron bound states (via strong positron-atom correlations)
was predicted in 1995 by many-body-theory calculations (for Mg, Zn, Cd and Hg) 
\cite{PhysRevA.52.4541}, and later confirmed by variational calculations for positronic lithium ($e^+$Li) \cite{PhysRevLett.79.4124,Strasburger}. 
Subsequently,  many more atoms and ions were studied \cite{Mitroy_2002}, and according to a comprehensive survey of the periodic table by \citet{Harabati14}, it is now expected that about 50 (ground-state) atoms can bind a positron. 
 Compared to the anions, which bind the positron in even the static level of theory, positron binding to neutral atoms 
relies on positron-atom correlation effects that overcome the short-range repulsive interaction with the nucleus \cite{Mitroy_2002,PhysScripta.46.248}. They make accurate calculations challenging.
Calculations have been performed using the stochastic variational method (SVM)
\cite{ECGRMP}, which is expected to be highly accurate, but it has been used only for systems with a maximum of five explicitly correlated quantum particles \cite{Bromley06}.  
Alternatively, the reduced explicitly correlated Hartree-Fock (RXCHF) method, based on the Hartree-Fock method employing explicitly correlated Gaussians, has produced some promising results, but it exhibits a strong dependence on the choice of correlation factor and the calculations so far have been limited to small systems, though they are expected to be more tractable than the corresponding SVM calculations \cite{Brorsen17}.  
In their survey of the periodic table, \citet{Harabati14} performed relativistic coupled-cluster (CC) calculations (treating core orbitals approximately) 
and found binding energies that were systematically larger than the SVM results, e.g., for Be, the simplest system considered, they found a binding energy of 214 meV compared to the SVM result of 86 meV.
The reason for the discrepancy was unclear, but they nevertheless use this 128 meV error for Be as an {\it ad hoc} correction for larger atoms \cite{Harabati14}.
The calculated data from \citet{Harabati14} have recently been analysed via a machine-learning approach by \citet{Amaral21} who developed an approximate formula for the positron-atom binding energy as a function of the atomic dipole polarizability $\alpha$ and ionization potential $I$ in the form $\varepsilon_b=A+B\exp(C\alpha^2 I)$, with fitting coefficients $A,B,C$ \cite{Amaral21}.
However, there is an apparent lack of consensus between the calculations, with distinct methods predicting binding energies differing by several hundreds of meV \cite{Harabati14}. 
This unsatisfactory situation is exacerbated by the fact that whilst experimental proposals to measure positron-atom binding energies exist \cite{Mitroy1999,PhysRevLett.105.203401,Surko2012,Swann2016}, none have yet been realized.

Recently, we developed a Gaussian-basis many-body theory approach that provided the first \emph{ab initio} calculated positron binding energies in agreement with experiment for several polar and non-polar molecules \cite{Jaro22}. 
We have since extended the method to halogenated hydrocarbons, highlighting the dependency of binding on molecular properties within molecular families \cite{Cassidy:2023}, and also to 
successfully describe positron scattering and annihilation on the small non-binding molecules H$_2$, N$_2$ and CH$_4$ \cite{Charlie23}, and examine the effects of high-order diagrams on positron scattering on noble-gas atoms \cite{JaroFrontiers2023}.
Here, we apply our many-body theory approach implemented in our {\tt EXCITON+} program \cite{Jaro22} to calculate positron binding energies  in the negative ions H$^-$, F$^-$, Cl$^-$ and Br$^-$, and the closed-shell atoms Be, Mg, Zn and Ca. 
We compare our results against other theoretical methods, including earlier atomic many-body theory (AMBT) calculations that employed a B-spline basis \cite{Ludlow10}.
To obtain accurate results for these neutral atoms, the interplay of attractive and repulsive interactions over both short and long range must be evaluated accurately.  
Notably, the polarization of the electron cloud by the positron,
screening of the electron-positron Coulomb interaction,  as well as the non-perturbative process of virtual-positronium (Ps) formation must be accounted for. 
The latter is especially important as the ionisation energies of these atoms are close to Ps formation energy $E_{\rm Ps}=$6.8~eV.
Compared with the earlier B-spline AMBT \cite{Ludlow10}, here we include higher-order diagrams, specifically calculating the electron-hole two-particle polarization propagator at the Bethe-Salpeter-Equation level (i.e., including the time-dependent Hartree Fock ring series with screened electron-hole intra-ring interactions) and the virtual-positronium and positron-hole ladder series with ladder rungs comprising of screened (as opposed to bare) Coulomb interactions.
Compared with the molecular systems, we find that the screening in the infinite ladder series diagrams plays a more significant role here. 

The outline of the remainder of the paper is as follows. In Section \ref{sec:theory} we give an overview of the many-body theory, followed in Section \ref{sec:numerics} by details of the numerical implementation via employment of Gaussian bases. In Sections \ref{sec:results-ions} {A} and {B} the calculated binding energies and annihilation rates are presented for negative ions and neutral-atoms respectively, before concluding with a summary.

\section{Theory}\label{sec:theory}
\begin{figure}[t!!]
\includegraphics*[width=0.48\textwidth]{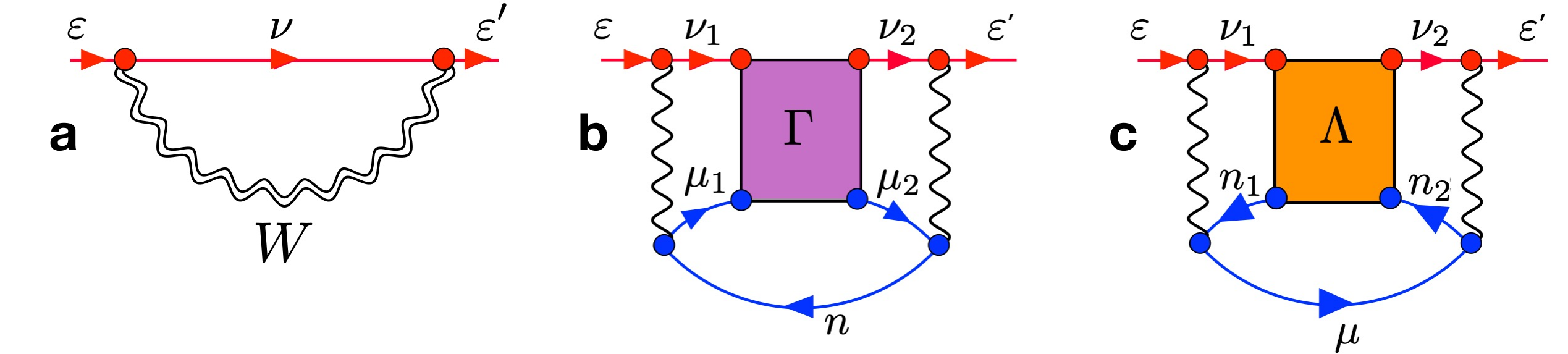}~~~
\caption{The main contributions to the positron-molecule self energy.
{\bf a}, the $GW$ diagram, which we calculate at the Bethe-Salpeter equation (BSE) level, describes the polarization of the molecular electron cloud and corrections due to screening of the electron-positron Coulomb interaction by the molecular electrons, and electron-hole attractions;
{\bf b} and {\bf c},  the summed infinite ladder series of screened electron-positron interactions (`$\Gamma$-block') and positron-hole interactions (`$\Lambda$-block').
\label{fig:diags}}
\end{figure}

In this section, we give an overview of our many-body theory \cite{Cederbaum-elecpos,Gribakin:2004,dzuba_mbt_noblegas} approach.  A more comprehensive description of the method is provided in \cite{Jaro22}.
Briefly, we calculate positron wave function $\psi_{\epsilon}$ and binding energy $\epsilon$ by solving the Dyson equation
\begin{equation}
\left(H^{(0)}+\hat{\Sigma}_{\epsilon}\right)\psi_{\epsilon}(\bm r) 
= \epsilon \psi_{\epsilon}(\bm r),\label{eqn:dyson}
\end{equation}
where $H^{(0)}$ is the zeroth order Hamiltonian for the positron in the field of the ground-state atom, 
and $\hat{\Sigma}_{\epsilon}$ is the positron-atom correlation potential (irreducible self-energy) whose energy dependence demands the equation be solved self-consistently. 
In practice, we compute $\hat{\Sigma}(E)$ at different energies $E$ close to $\epsilon$ and report the lowest eigenvalue, equal to the negative of the binding energy,  at the intercept $\epsilon=E$.
We expand the self-energy in residual electron-electron and electron-positron interactions as shown in Figure~\ref{fig:diags} (see also Figure~1 in \cite{Jaro22} for more details).
The first diagram we include is the $GW$ diagram, the simplest approximation to which we consider is the second-order polarization diagram $\Sigma^{(2)}$ that describes the bare polarization of the electron cloud by the positron.
Beyond this, we also consider screening at the RPA (Random Phase Approximation) level that
includes the infinite electron-hole `bubble' or `ring' series of diagrams on top of $\Sigma^{(2)}$, 
 and also TDHF (Time Dependent Hartree-Fock) and BSE (Bethe-Salpeter Equation), which respectively
 additionally include include bare or dressed electron-hole Coulomb interactions within the `bubbles'
Further, we take into account the virtual-positronium (Ps) formation, which takes the form of an infinite ladder series of electron-positron interactions, causing positron attraction to the electron cloud.
This is represented by the shaded $\Gamma$-block in the diagrams, and denoted by $\Sigma^{\Gamma}$.
Finally, we also consider the (screened) infinite series of positron-hole repulsive interactions $\Lambda$, which is similar to $\Gamma$ in structure (see Figure~\ref{fig:diags}).
We calculate the total self-energy as $\Sigma=\Sigma^{GW}+\Sigma^{\Gamma}+\Sigma^{\Lambda}$.

Earlier atomic many-body theory (AMBT) calculations used a B-spline basis to calculate binding energies for the negative ions \cite{Ludlow10}.
AMBT treated electron screening beyond third-order diagrams approximately.
In contrast, here we implement the $GW$ diagram at the BSE level, which includes the RPA polarization (including screened electron-hole interactions), and we also calculate the virtual-Ps and positron-hole ladder contributions using dressed Coulomb interactions. 
Thus, we can directly benchmark our present Gaussian-basis approach at Hartree-Fock,  bare polarization $\Sigma^{(2)}$, and $\Sigma^{(2+\Gamma)}$ levels of theory \cite{Ludlow10}.

Solution of the Dyson equation (Equation~\ref{eqn:dyson}) yields not only the positron binding energy but also the positron-bound state wave function $\psi_{\varepsilon}$. 
Using it, the 2$\gamma$ annihilation rate in the bound state $\Gamma= \pi r_0^2c\delta_{ep}$ ($\Gamma[{\rm ns}^{-1}]= 50.47\,\delta_{ep}[{\rm a.u.}])$, whose inverse is the lifetime of the positron-atom complex with respect to annihilation, can be calculated. 
Here $r_0$ is the classical electron radius, $c$ is the speed of light and $\delta_{ep}$ is the contact density, which reads in the independent particle approximation
\begin{equation}\label{eqn:delta_ep}
\delta_{ep}=\sum_{n=1}^{N_e}  \gamma_{n} \int   |\varphi_n({\bm r})|^2 |\psi_{\varepsilon}({\bm r})|^2d {\bm r},
\end{equation}
where $\gamma_n$ are orbital dependent enhancement factors that account for the short-range electron-positron attraction \cite{DGG:2015:core,DGG:2017:ef}.
Previous many-body calculations for atoms determined the enhancement factors to follow a physically motivated scaling with the orbital energy $\varepsilon_n$ in atomic units \cite{DGG:2015:core,DGG:2017:ef}  
\begin{equation}\label{eqn:gamma}
\gamma_n = 1+ \sqrt{ {1.31}/{|\varepsilon_n|}} + \left({0.834}/{|\varepsilon_n|} \right)^{2.15}.
\end{equation}
We also report unenhanced contact densities $\delta^{(0)}_{ep}$ with the enhancement factor $\gamma_n=1$.

The positron Dyson wave function is a quasi-particle wave function that is the overlap of the wave function of the $N$-electron ground state atom with the fully-correlated wave function of the positron plus $N$-electron atom system \cite{mbtexposed}. 
It is normalised as 
\begin{eqnarray}\label{eqn:aval}
\int |\psi_{\varepsilon}({\bm r})|^2 d{\bf r}= \left(1-{\partial \varepsilon}/{\partial E}\right|_{\varepsilon_b})
^{-1} \equiv a <1,
\end{eqnarray} 
which estimates the degree to which the positron-atom bound state is a single-particle state, with smaller values of $a$ signifying a more strongly correlated state. 
For the negative ions as well as Ca,  the ionization energy $I<E_{\rm Ps}=6.8$~eV, meaning that the lowest dissociation threshold in the positron-anion system is the neutral atom A and positronium (PsA),  and consequently, for the e$^+$A$^-$ system to be bound,  the positron binding energy should be higher than $E_{\rm Ps}-I$ (see Ref.~\cite{Mitroy_2002} for more details).

\section{Numerical implementation}\label{sec:numerics}
We implement the above in the massively-parallelized {\tt EXCITON+} program developed by us \cite{Jaro22}.
We work with the matrix elements of the self energy in the Hartree-Fock basis, 
expanding the electron (--) and positron (+) Hartree-Fock orbitals in distinct Gaussian basis sets as 
\begin{equation}
\varphi_{a}^{\pm}(\bm{r})=\sum_{A}^{N_c^{\pm}}\sum_{k=1}^{N_A^{\pm}} C_{a Ak}^{\pm} \chi^{\pm}_{A_k}(\bm{r}),
\end{equation}
where $A$ labels the $N_c^{\pm}$ basis centres, $k$ labels the $N_A^{\pm}$ different Gaussians on centre $A$, each taken to be of Cartesian type with angular momentum $l^x+l^y+l^z$, such that
\begin{equation}
\chi_{A_k}(\boldsymbol{r}) = \mathcal{N}_{A_k}(x-x_A)^{l^x_{Ak}}(y-y_A)^{l^y_{Ak}}(z-z_A)^{l^z_{Ak}} e^{-\zeta_{Ak} |{\bf r-r}_A|^2},
\end{equation}
where $\mathcal{N}_{A_k}$ is a normalisation constant, and $C$ are the expansion coefficients to be determined by solving the Hartree-Fock (Roothaan) equations.
Expanding the positron Dyson wave function (see Equation~\ref{eqn:dyson}) in the positron HF MO basis as $\psi_{\varepsilon}({\bf r})= \sum_{\nu} D^{\varepsilon}_{\nu} \varphi^+_{\nu}({\bf r})$ transforms the Dyson equation to the linear matrix equation $\bm{HD}=\bm{\varepsilon D}$, where $ \langle \nu_1 | H | \nu_2\rangle = \varepsilon_{\nu_1} \delta_{\nu_1\nu_2} + \langle \nu_1 |\Sigma_{\varepsilon} | {\nu_2}\rangle$. 
We construct the individual contributions to $\Sigma_{\varepsilon}$ by first solving the respective Bethe-Salpeter equations for the electron-hole polarization propagator $\Pi$, the two-particle positron-electron propagator $G^{\rm ep}_{\rm II}$ and the positron-hole two-`particle' propagator $G_{\rm II}^{\rm ph}$\cite{mbtexposed,Jaro22}. 

For the electronic wave function expansion, we use Dunning's correlation consistent basis sets with augmentation functions \cite{Dunning} denoted aug-cc-pVXZ, where X is T or Q. 
For the positron wave function description, we use a diffuse even-tempered Gaussian basis such that for functions of specific angular-momentum type $l$ (e.g. $s,p,d$) centred on nucleus A, we choose the exponents $\zeta_{Aj}$ as
$\zeta_{Aj}=\zeta_{A1}\beta^{j-1} \quad (j = 1, \dots , N_A^{l})$, where $\zeta_{A1}>0$ and $\beta>1$ are parameters. 
The value of $\zeta_{A_1}$ is important as the bound positron wavefunction behaves asymptotically as $\psi\propto e^{-\kappa r}$, where $\kappa=\sqrt{2\varepsilon_b}$. 
Thus, to ensure that the expansion describes the wave function well at $r\sim 1/\kappa$, i.e., that the broadest Gaussian covers the extent of the positron wave function, one must have $\zeta_{A_1}\lesssim \kappa^2=2\varepsilon_b$.
In practice we performed binding energy calculations for a range of $\zeta_{A_1}$ and $\beta$ for each system, finding that there are broad ranges of stability.  
The optimal $\zeta_{A_1}$ was found to be $10^{-3}$ or $10^{-2}$ for $s$- and $p$-type Gaussians and $10^{-2}$ or $10^{-1}$ for $d$-, $f$- and $g$-type Gaussians, with the former (latter) values valid for weakly binding neutrals (strongly binding ions). 
The optimal $\beta$ was found to be equal to 2.2 for the basis set $10s9p8d7f6g$ used predominantly in this work.

We place additional (hydrogen type) aug-cc-pVQZ basis sets on `ghost' centres for both electrons and positrons that generate effectively higher angular momenta basis functions \cite{Swann:2018} and increase the flexibility and completeness of the positron and electron bases, thus aiding convergence. 
Their locations are manually optimized. 
Note that we do not perform extrapolation with respect to the basis set angular momenta, as is common in other similar calculations, such as B-spline many-body theory \cite{Gribakin:2004,Ludlow10} or Configuration Interaction \cite{Bromley06}.
\section{Results and discussion}\label{sec:results-ions}
\subsection{Calculated static dipole polarizabilities and ionization energies}
The calculated static dipole polarizabilities $\alpha$ and ionization potentials $I$ of the studied atoms are shown in Table~\ref{tab:ion_pol} along with calculated reference data, compared to which our values are typically smaller. 
The largest dipole polarizability of all atoms is computed for H$^-$ and it is found to converge very slowly with the basis set size and maximum angular momentum: additional basis sets on ghost centres would likely improve agreement with the reference calculation \cite{Polar2020}. 
Nevertheless, we find binding energies in agreement with other theory (see below).   
According to the present calculations, HF polarizabilities are always lower than the BSE ones, with the notable cases of Be, Zn, and Ca atoms,  for which the BSE polarizabilities are higher than the reference data.

\begin{table}[!t]
\caption{\label{tab:ion_pol}Static dipole polarizabilities (in \AA$^3$) and ionization energies (in eV) of studied atoms calculated using aug-cc-pVQZ basis sets and including additional ghost centres located on a sphere about 1~\AA~ away.  
}
\begin{ruledtabular}
\begin{tabular}{l@{\hskip 18pt}c@{\hskip 7pt}c@{\hskip 7pt}c@{\hskip 7pt}c@{\hskip 7pt}c@{\hskip 7pt}c@{\hskip 7pt}c}
&\multicolumn{4}{c}{Isotropic polarizability $\bar{\alpha}$ (\AA$^3$) }&\multicolumn{3}{c}{Ionization energy (eV)}\\
 \cline{2-5} \cline{6-8}
& HF 
&RPA
&BSE  
& Ref. \footnote{Reference values are from \cite{CRC97}, except for the polarizabilities of ions, which are taken from \cite{Polar2020}. } 
& HF 
& $GW$
\footnote{Ionisation energies calculated at the $GW$@RPA level were obtained using the diagonal approximation for the electron-atom self energy  
$\Sigma^{(-)}$
\cite{Reiningbook}.} & Ref.\footnotemark[1]\\
H$^-$ &6.08& 4.97&12.7&34.0 & 1.24 & 1.05 & 0.75\\
F$^-$ &1.11& 0.92&1.95&2.46 & 4.92 & 2.97 & 3.40\\
Cl$^-$ &3.78& 2.69&4.88&5.36 & 4.09 & 3.72 & 3.61\\
Br$^-$ &5.25& 3.62&6.37&7.26 & 3.79 & 3.59 & 3.36\\[0.8ex] 
Be &4.53& 2.99&5.94&5.60 & 8.42 & 9.14 & 9.32\\
Mg &8.20& 5.35&10.5&10.6 & 6.88 & 7.60 & 7.65\\
Zn &5.82& 3.72&6.85&5.75 & 7.96 & 8.88 & 9.39\\
Ca &23.3&13.9&28.6&25.0 & 5.04 & 5.81 & 6.11\\
\end{tabular}
\end{ruledtabular}
\end{table}%

\subsection{Positron binding energies in negative ions}
Table \ref{tab:halides} shows our calculated binding energies for the negative ions. 
We first benchmark our approach by considering positron binding energies for Cl$^-$: for this ion there are detailed atomic many-body theory (AMBT) calculations \cite{Ludlow10} reported at different approximations to the self-energy diagrams to which we can directly compare. 
At the static HF level,  we calculate the positron-Cl$^-$ binding energy of 3.86 eV (0.1419 a.u.), in perfect agreement with AMBT \cite{Ludlow10}.  
Even though the HF approximation provides a strongly bound state already,  adding the self-energy on top of HF is essential in determining an accurate binding energy. 
Starting with the second-order $GW$ polarization diagram,  at $\Sigma^{(2)}$ bare polarization level we obtain 5.03 eV (compared to 5.05 eV of AMBT) in near-perfect agreement.
Next, the inclusion of the virtual Ps contribution increases binding by further 1 eV (20\%),  such that at $\Sigma^{2+\Gamma}$ level, our method gives a converged binding energy of 6.10~eV, again in a very close agreement with AMBT's 6.19 eV (about 1.5\% discrepancy).
However, the inclusion of the ghost centres generating effectively high angular momenta is important to provide good agreement as the convergence is relatively slow: without ghost centres, the binding energy at this level of theory is 5.77~eV.
We used up to 20 ghost centres, placed at vertices of a regular dodecahedron at $1.1 - 1.6$~\AA\, distance from the atom at the origin (optimized for each species), with aug-cc-pVQZ basis sets of H.  
The use of ghost centres increased binding energies by as much as 3\% in the case of $GW$ approximation and about 5\% for the virtual-Ps results. 
We note that the AMBT extrapolates binding energies to infinite angular momenta: at $\Sigma^{2+\Gamma}$ level this changes the binding energy by 0.5~\% \cite{Ludlow10}.
Our method effectively includes high angular momenta via the use of ghost centres, without the need to extrapolate.

\begin{figure}[!t]
\adjustbox{trim={.02\width} {.04\height} {0.15\width} {.05\height},clip}{\includegraphics*[scale=0.35]{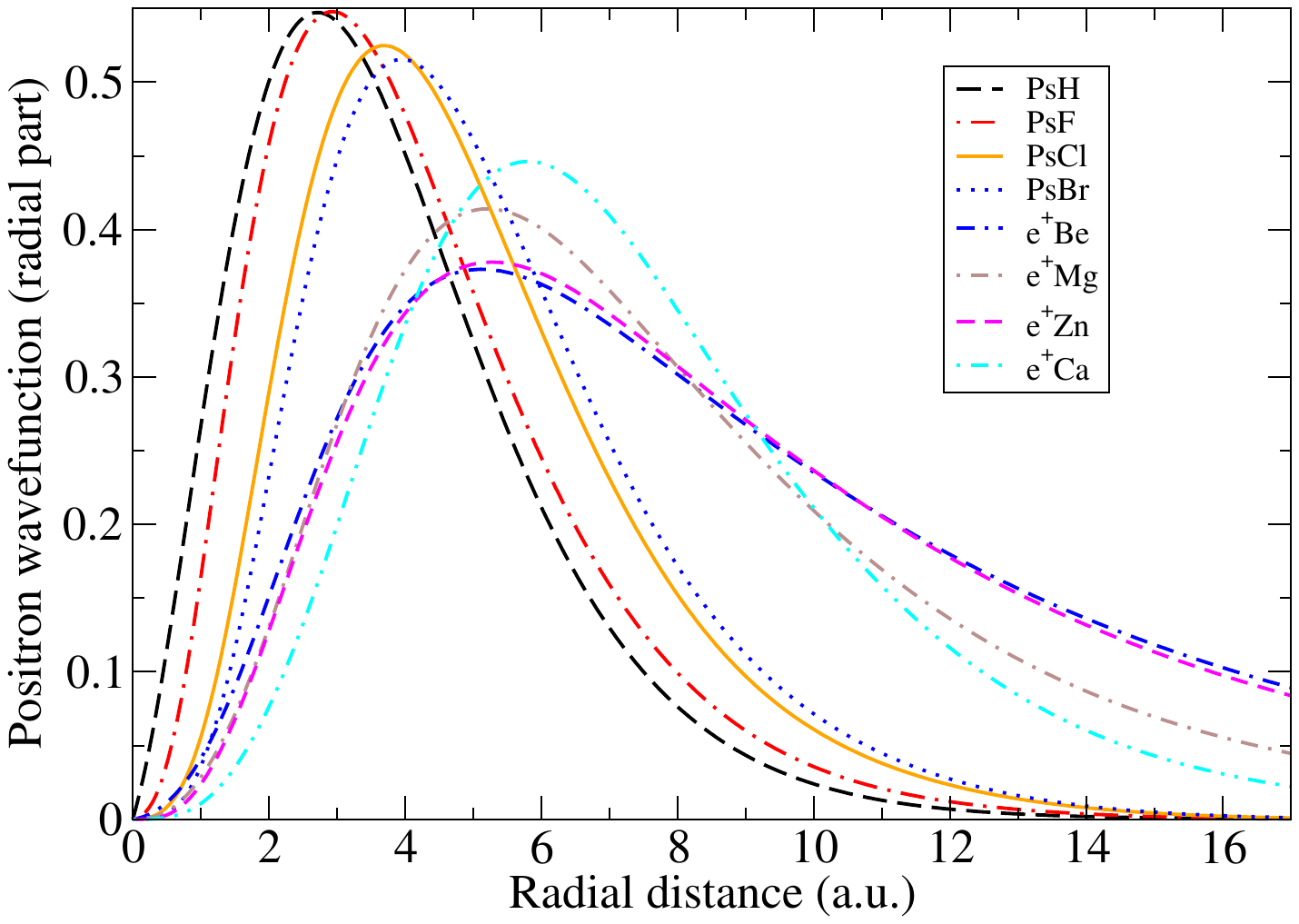}}
\caption{\label{radwfnoa} The positron bound state wavefunction (radial part) along the radial axis for the systems studied in this work obtained from the Dyson equation with $\Sigma^{GW+\tilde{\Gamma}+\tilde{\Lambda}}$. 
Note that the wave functions are normalised to unity for comparison purposes.} 
\end{figure}

Of the studied ions, the largest Hartree-Fock binding energy is found for F$^-$ (about 5~eV, agreeing with an older result of \citet{Patrick81}),  and smallest for Cl$^-$ and Br$^-$ due to stronger positron repulsion from the positively-charged atomic cores in these systems (see Table~\ref{tab:halides}).
Starting with the bare polarization $\Sigma^{(2)}$ approximation,  the first positron bound state is strongest in H$^-$ and weakens through the halogen sequence to Br$^-$.
The inclusion of BSE electron-hole pair screening in the $GW$ diagram reduces the $\Sigma^{(2)}$ binding energies for Cl$^-$ and Br$^-$,  but increases them for H$^-$ and F$^-$ (see Table~\ref{tab:halides}).
Thus, there is a delicate balance of attractive polarisation, electron-hole screening via the random phase approximation (RPA),  and intra-ring electron-hole attractive corrections to the screening.
Further, the positron-hole ladder series contribution $\Sigma^{\Lambda}$ lowers the $\Sigma^{GW+\Gamma}$ binding energies slightly, ranging from 0.22~eV or about 3\% in F$^-$ to 0.36 eV or 5\% in H$^-$.
Finally, including (RPA-type) screening in both ladder series diagrams results in our most sophisticated $\Sigma^{GW+\tilde{\Gamma}+\tilde{\Lambda}}$ binding energies,  reported as two values, one using HF orbital energies and the other using $GW$ energies in the energy denominators in the diagrams (see \cite{Jaro22} for the diagram expressions). 
The latter is close to the $\Sigma^{GW+\Gamma+\Lambda}$ result.
Overall, the positron binding energies decrease as the atom size grows due to a stronger repulsion from the positively-charged nuclei (with the exception of HF results for H$^-$ and F$^-$).
Moreover, as expected, radial parts of the positron wave function peak further away from the nucleus (see Figure~\ref{radwfnoa}).
Comparing our best results to previous AMBT, MRCI and DMC calculations (in Table~\ref{tab:halides}), which are in very good mutual agreement \cite{Ludlow10,Saito03,Saito05,bressanini98,Ito2020,Charry2022},  we can see that our most sophisticated results are usually higher by 2--6\%, with the higher end of the range being for H$^-$. 
This agreement is however not entirely satisfactory. We also find typically larger binding energies in the weakly bound systems (see the next section).
Notably,  our results match quite well with AMBT's best \emph{ab initio} $\Sigma^{(2+3+\Gamma)}$ results.

Values of the normalization parameter $a$ (see Equation~\ref{eqn:aval}) quantify the strength of correlations within the positron-anion system.  
F$^-$ has the largest value of $a=0.923$ with the values being slightly lower for Cl$^-$ ($a=0.865$), Br$^-$ ($a=0.838$), and lowest for H$^-$ ($a=0.70$).
This indicates that a positron bound to the negative ion is the dominant component of the structure \cite{Mitroy_2002}.  
We note that these results are in very good agreement with the AMBT results \cite{Ludlow10} at $\Sigma^{(2+3+\Gamma)}$ level of theory, namely 0.714 (H$^-$), 0.950 (F$^-$) , 0.875 (Cl$^-$), and 0.834 (Br$^-$). 
The contact densities obtained at the $\Sigma^{GW+\tilde{\Gamma}+\tilde{\Lambda}}$ approximation are also in good agreement with previous AMBT results \cite{Ludlow10} for Cl$^-$.
The presently calculated value of $\delta^{(0)}_{ep}=8.61\times10^{-3}$~a.u. lies in between the fully correlated $\Sigma^{\rm scr}$ result of $8.72\times10^{-3}$~a.u. and the best \emph{ab initio} $\Sigma^{2+3+\Gamma}$ result of $8.41\times10^{-3}$~a.u. \cite{Ludlow10}.
The enhancement factors $\gamma_n$ given by Equation~\ref{eqn:gamma} are unrealistically large due to the final term, which does not behave well for negative ions with small $I$. 
Without the final term, the enhancement factor for Cl$^-$ is about 4.1, which is reasonably close to 4.5 determined diagrammatically within AMBT \cite{Ludlow10}.
To avoid using the approximate enhancement factors, a calculation of the annihilation vertex corrections using the multi-centred Gaussian basis approach is needed, but is computationally extremely challenging, and substantial developments are required that are beyond the scope of this work. 

\begin{table*}[t!]
\caption{Calculated positron binding energies for negative ions (eV). 
The present many-body theory calculated binding energy calculations are presented in the Hartree-Fock,  $\Sigma^{(2)}$ (bare polarization) and $GW$@BSE (bare polarization including dressed electron-hole interactions) approximations, 
and additionally including virtual-Ps formation $\Sigma^{GW+\Gamma}$, 
and the positron-hole ladder contribution $\Sigma^{GW+\Gamma+\Lambda}$.  Our most sophisticated approximation $\Sigma^{GW+\tilde{\Gamma}+\tilde{\Lambda}}$ uses dressed Coulomb interactions in the $\Gamma$ and $\Lambda$ blocks: for this we report two values, obtained using either HF or $GW$ energies in the diagrams (the latter the recommended best, in bold).
\label{tab:halides}}
\begin{center}
\begin{ruledtabular}
\begin{tabular}{l@{\hskip8pt}c@{\hskip8pt} ccccc@{\hskip9pt} ccccc}
 & &\multicolumn{5}{c}{Present many-body theory}  & \multicolumn{5}{c}{Other calculations}\\
 \cline{3-7} \cline{8-12}
  & &\multicolumn{5}{c}{ }  & \multicolumn{2}{c}{AMBT\footnote{Previous B-spline based atomic many-body theory calculations included second-order and third-order Goldstone diagrams and virtual-Ps in $\Sigma^{2+3+\Gamma}$, and additionally took approximate account of higher order screening diagrams in $\Sigma^{\rm scr}$ \cite{Ludlow10}.}}&APMO\footnote{APMO REN-PP3 method \cite{Romero:2014,APMO2020}.}& MRCI\footnote{Multi-reference Configuration Interaction \cite{Saito03,Saito05}.} & DMC\footnote{Diffusion Monte Carlo: \cite{Ito2020,Charry2022} (H$^{-}$) and \cite{bressanini98} (F$^{-}$).}\\
			& HF		& $\Sigma^{(2)}$ 		& $\Sigma^{GW}$ 	& $\Sigma^{GW+\Gamma}$ 	& $\Sigma^{GW+\Gamma+\Lambda}$ & $\Sigma^{GW+\tilde{\Gamma}+\tilde{\Lambda}}$	& $\Sigma^{2+3+\Gamma}$ & $\Sigma^{\rm scr}$  & & &  \\[1ex]
H$^{-}$	&	4.50		&	5.96		&	6.06	& 	 7.91	& 7.55	&7.31, {\bf 7.56}	& 7.52 & 7.17 &  6.70 & 7.11 & 7.11	\\[0.5ex]
F$^{-}$		&4.96		&5.93		& 6.01	& 6.53 	&6.31	&6.27, {\bf 6.35}& 6.20 &6.12 & 5.95 &6.23	& 6.17	\\[0.5ex]
Cl$^{-}$	&3.86		&5.03		& 4.98 & 6.00 	&5.73	&5.63, {\bf 5.72}	&  5.63  &5.44 & 5.15 &5.52	 & --	\\[0.5ex]
Br$^{-}$ 	  &3.59	&4.79		& 4.71	&  5.85    & 	5.57    &5.45, {\bf 5.56} &  5.55 &5.31 & 4.89 &5.36	& --	\\
\end{tabular}
\end{ruledtabular}
\end{center}
\end{table*}%

\subsection{Closed shell neutral atoms}
Table \ref{tab:atoms} shows the calculated positron binding energies for the atoms Be, Mg, Zn and Ca.
Neutral atoms do not bind the positron at the Hartree-Fock level of theory \cite{Patrick81}.
At the $GW$ level, Be,  Mg, and Zn (with $I>6.8$ eV, see Table~\ref{tab:ion_pol}) still do not support a positron bound state. 
On the other hand,  Ca (with $I<6.8 $ eV) binds the positron at $GW$ level of theory though very weakly.  
The presently calculated ionization energies at HF and $GW$ levels of theory are smaller in magnitude than the experimental ionization energies (see Table~\ref{tab:ion_pol}) and moreover,  the BSE polarizabilities are higher than the reference values (except for Mg),  indicating the present method might be expected to overestimate binding owing to the greater ease of the positron interacting with relatively more diffuse bound electrons, and a stronger polarisation potential attraction. 
\begin{table}[h!]
\caption{Calculated positron-atom binding energies (meV), bound state contact densities without enhancement $\delta^{(0)}_{ep}$,  and enhanced $\delta_{ep}$ (with $\gamma_n$ in Eq.~\ref{eqn:gamma}).
The results are presented at $GW$@BSE (bare polarization plus screening and electron-hole corrections) level, with virtual-positronium formation $\Sigma^{GW+\Gamma}$ level of theory, 
and including the positron-hole ladder contribution $\Sigma^{GW+\Gamma+\Lambda}$. Our best approximation $\Sigma^{GW+\tilde{\Gamma}+\tilde{\Lambda}}$ uses dressed Coulomb interactions in the $\Gamma$ and $\Lambda$ blocks and we use $GW$ energies in the diagrams instead of SCF.
Numbers in brackets indicate powers of 10.
The $p-$type bound states are in the second row,  wherever present, marked with an asterisk.
\label{tab:atoms}}
\begin{center}
\begin{ruledtabular}
\begin{tabular}{l@{\hskip4pt}c@{\hskip5pt}c@{\hskip5pt} c@{\hskip5pt}c}
Calculation& $\epsilon_b$ (meV) & $\delta^{(0)}_{ep}$ (a.u.)& $\delta_{ep}$ (a.u.) & $a$ \\[1.5ex]
\hline\\[0.5ex]
\multicolumn{5}{c}{\textbf{Be}} \\
Present MBT, $\Sigma^{GW}$ & $<0$ & -- &  -- & -- \\
Present MBT, $\Sigma^{GW+\Gamma}$ & 550 & 1.16[$-$3] & 1.08[$-$2] & 0.77 \\
Present MBT, $\Sigma^{GW+\Gamma+\Lambda}$ & 420 & 1.02[$-$3] &  9.42[$-$3] & 0.79 \\
Present MBT, $\Sigma^{GW+\tilde{\Gamma}+\tilde{\Lambda}}$ 	& {\bf 290} & 0.87[$-$3] &  8.20[$-$3] & 0.82 \\[1.5ex]
Relat. coupled cluster \cite{Harabati14} & 214 &   &    &   \\
Stochastic variational \cite{Mitroy10} & 86 &  &  8.55[$-$3] &  \\[1.5ex]
\multicolumn{5}{c}{\textbf{Mg}} \\[0.5ex]
Present MBT, $\Sigma^{GW}$ & $<0$ & -- &  -- & -- \\
Present MBT, $\Sigma^{GW+\Gamma}$ & 1065 & 1.07[$-$3] &  1.16[$-$2] & 0.62 \\
& 195$^\ast$& 4.63[$-$4] &  5.70[$-$2] & 0.64 \\
Present MBT, $\Sigma^{GW+\Gamma+\Lambda}$ & 935 & 1.01[$-$3] &  1.09[$-$2] & 0.63 \\
& 114$^\ast$  & 4.19[$-$4] & 5.18[$-$3] & 0.65 \\
Present MBT, $\Sigma^{GW+\tilde{\Gamma}+\tilde{\Lambda}}$ 	& {\bf 710} & 0.93[$-$3] & 1.02[$-$2] & 0.69 \\[1.5ex]
Earlier MBT \cite{gribakin96}   & 985 &   &    &   \\
  & 159$^\ast$ &   &    &   \\
Relat. coupled cluster \cite{Harabati14} & 636 &   &   &  \\
Config. Interaction \cite{Bromley06} & 464 &   &  1.96[$-$2] &  \\[1.5ex]
\multicolumn{5}{c}{\textbf{Zn}} \\[0.5ex]
Present MBT, $\Sigma^{GW}$ & $<0$ & -- &  -- & -- \\
Present MBT, $\Sigma^{GW+\Gamma}$ & 645 &   1.82[$-$3] &  1.29[$-$2] & 0.76 \\
Present MBT, $\Sigma^{GW+\Gamma+\Lambda}$ & 493 & 1.58[$-$3] &  1.10[$-$2] & 0.77 \\
Present MBT, $\Sigma^{GW+\tilde{\Gamma}+\tilde{\Lambda}}$ 	& {\bf 315} & 1.34[$-$3] & 9.76[$-$3] & 0.82 \\[1.5ex] 
Relat. coupled cluster \cite{Harabati14} & 235 &   &    &  \\
Config. Interaction \cite{Mitroy08} & 103 &   &  8.94[$-$3] &  \\[1.5ex]
\multicolumn{5}{c}{\textbf{Ca}} \\[0.5ex]
Present MBT, $\Sigma^{GW}$ & 36 & 1.19[$-$4] &  2.19[$-$3] & 0.95 \\
Present MBT, $\Sigma^{GW+\Gamma}$ & 2120  & 5.92[$-$4] &  9.56[$-$3] & 0.42 \\
& 1491$^\ast$ & 3.59[$-$4] &  6.67[$-$3] & 0.37 \\
Present MBT, $\Sigma^{GW+\Gamma+\Lambda}$ & 2015  & 5.67[$-$4] &  9.30[$-$3] & 0.42 \\
& 1415$^\ast$ & 3.48[$-$4] &  6.38[$-$3] & 0.37\\
Present MBT, $\Sigma^{GW+\tilde{\Gamma}+\tilde{\Lambda}}$ 	& {\bf 1590}  & 5.59[$-$4] & 9.37[$-$3] & 0.47 \\
& {\bf 1004}$^\ast$ & 3.50[$-$4] &  6.51[$-$3] & 0.44 \\[1.5ex] 
Relat. coupled cluster \cite{Harabati14} & 1432 &   &    &  \\
Config. Interaction \cite{Bromley06} & 1201 &   &  2.93[$-$2] &  \\[0.5ex]
\end{tabular}
\end{ruledtabular}
\end{center}
\end{table}%

\begin{figure}[ht!]
\adjustbox{trim={.02\width} {.02\height} {0.15\width} {.15\height},clip}{\includegraphics*[scale=0.33]{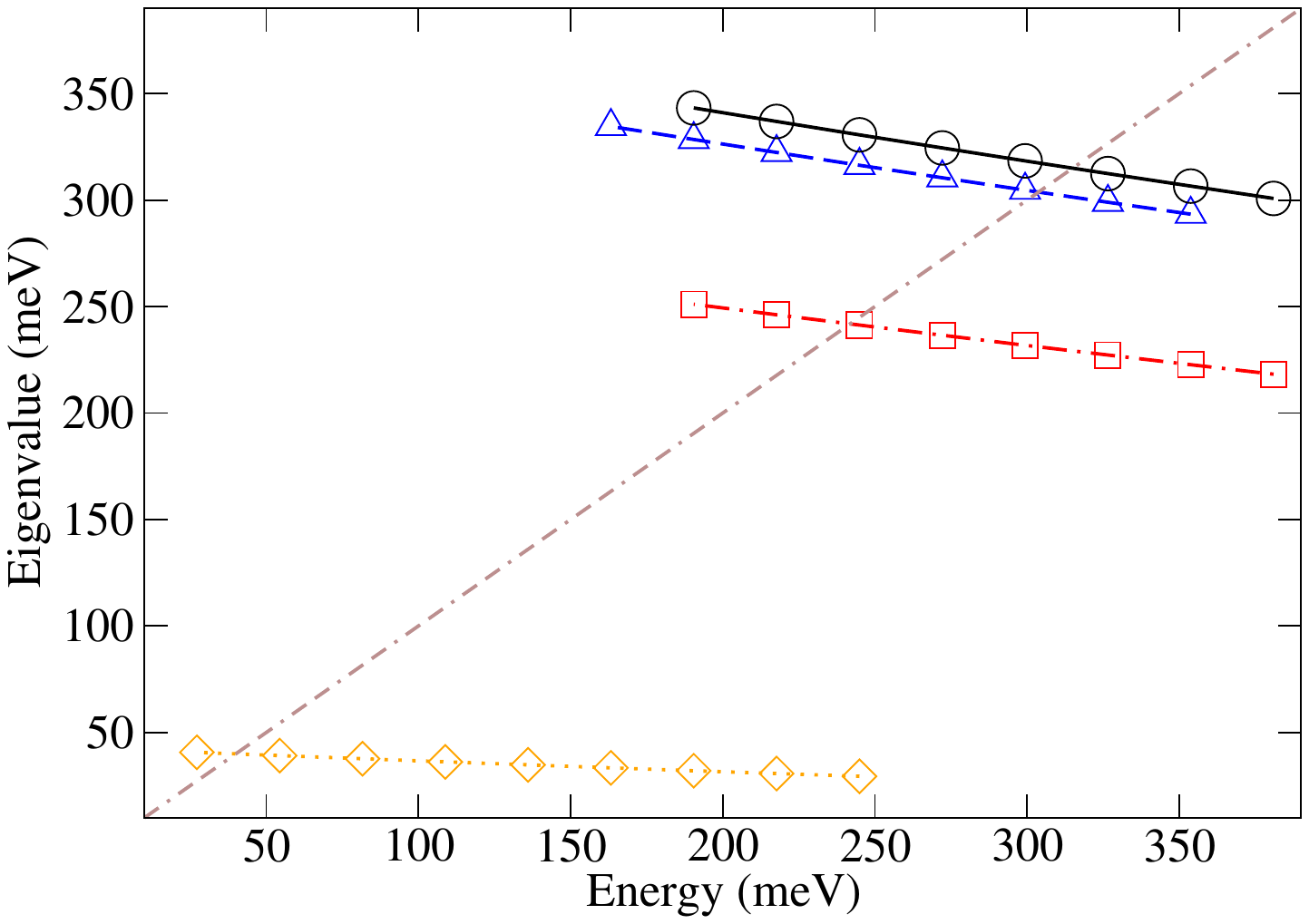}}\\[2ex]
\adjustbox{trim={.02\width} {.04\height} {0.15\width} {.05\height},clip}{\includegraphics[scale=0.33]{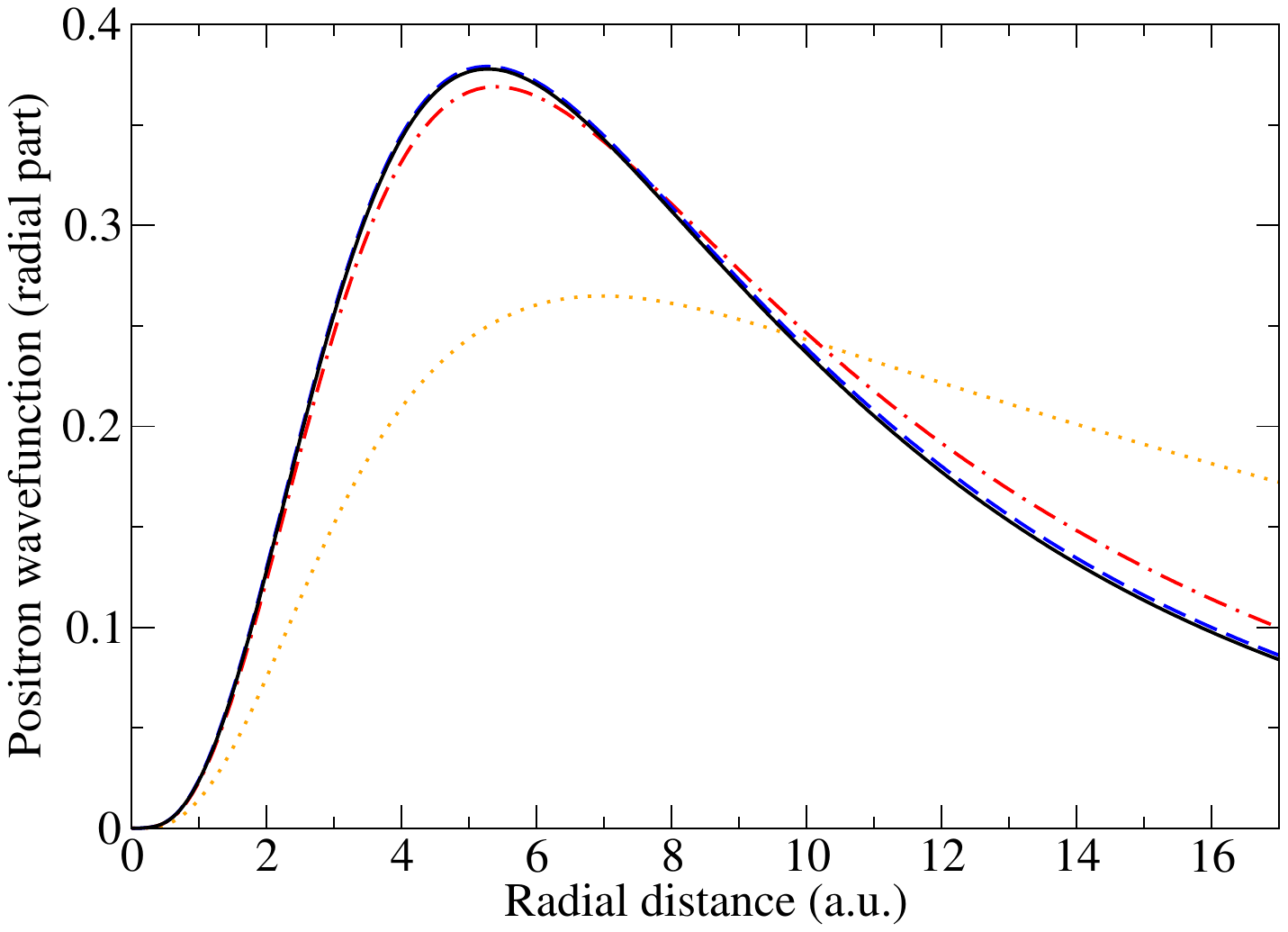}}
\caption{\label{evsE}Convergence of the positron bound state energy $\varepsilon_b(E)$ (top) and bound state radial wave function (bottom) for Zn computed using $\Sigma^{GW+\tilde{\Gamma}+\tilde{\Lambda}}(E)$. 
Different symbols refer to different basis sets used; diamonds (dotted lines): no ghost centres,  squares (dash-dotted lines): with 6 ghost centres,  triangles (dashed lines): with 12 ghost centres, and circles (solid lines): with 20 ghost centres.  
In the top plot, the dot-dash-dashed line is $\varepsilon = E$ and the intersection of the two lines, $\varepsilon_b(E) = E$, gives the binding energy. 
The gradient of $\varepsilon(E)$ at this point is used to calculate the normalization constant $a$ from Equation \ref{eqn:aval}. 
The basis sets on Zn are aug-cc-pVTZ (for e$^-$) and $10s9p8d7f6g$ (for e$^+$) while the ghost centres have aug-cc-pVQZ of H for both e$^-$ and e$^+$.  
 } 
\end{figure}

The inclusion of the virtual-positronium diagram $\Gamma$ is crucial to enable binding for Be, Mg, Zn atoms (see Table~\ref{tab:atoms}) and increases binding energy of the $s$-type bound state in Ca by a factor of 50.  
We note that earlier AMBT of Gribakin and King \cite{gribakin96} predicted the binding energy of $s$- and $p-$type bound states in Mg  as 985 meV and 159 meV, respectively, when the polarization diagram is taken into account together with an approximate treatment of virtual-Ps formation (where only the Ps ground state is considered).  
This is in good agreement with our $\Sigma^{GW+\Gamma}$ results (1065 meV and 190 meV) and $\Sigma^{GW+\Gamma+\Lambda}$ results (935 meV and 102 meV).  
Overall, the subsequent inclusion of positron-hole ladder series $\Lambda$ brings binding energies down by 10--20\% while the inclusion of screening within the diagrams is even more important, lowering the binding energies by 25--35\% and bringing them more in line with previous theoretical results. 
Interestingly, this contrasts with the behaviour seen for the negative ions,  where the positron-hole ladder series contribution is larger than that of screening.
Similarly to the negative ions, the convergence of the binding energies with respect to the basis set size is again quite slow.
Using 20 ghost centres enhanced the binding energies calculated without any ghost centres by as much as 40\% for Be and Mg and 80\% for Zn and Ca (see Figure~\ref{evsE}).
It is expected that further optimization of the basis sets could increase binding energies by $<$10\%.

The present results are consistently higher by 70--160 meV (or 15--25\%) than those of the \emph{ab initio} relativistic linearized coupled-cluster (CC) method \cite{Harabati14}, which reported accuracy of about 100 meV, determined presumably by comparing with an earlier stochastic variational method result \cite{Harabati14}.  
Our binding energies are higher than the recommended \cite{Swann:2020} reference results obtained by Mitroy and co-workers using explicitly correlated Gaussians (ECGs), stochastic variational methods (SVM) or Configuration Interaction (CI) methods \cite{Mitroy08,Bromley06,Mitroy10}.  
The most accurate reference calculation was done for Be \cite{Mitroy10} and our result is 190 meV above this reference, and 62 meV above the CC result.  
This could be due to the near-degeneracy of Be $2s$ and $2p$ states rendering the starting Hartree-Fock description inadequate \cite{Hollett2011}. However, we do not observe a significant overestimation of the dipole polarizability of Be, which would be expected in that case.
More sophisticated calculations not relying on the Hartree-Fock reference state may shed light on this discrepancy. 

The corresponding bound state contact densities are shown in Table~\ref{tab:atoms}.  
The zeroth-order (unenhanced) contact densities are highest for Zn, even though the binding energies are larger for Mg and Ca at all levels of theory. 
The enhancement factors (see Equation~\ref{eqn:gamma}) and renormalization constants together account for approximately a factor of 7 (8) increase in the contact densities for Be, Mg (Zn, Ca).
The agreement with the reference contact density results for Be and Zn is very good (within 5-10\%), but the results for Mg and Ca are surprisingly low. 
This could be caused by a slower convergence with respect to the maximum angular momenta compared to binding energies \cite{Bromley06,Ludlow10} or, alternatively,  the appearance of a $p$-type bound state (resonance) in Ca (Mg) might play a role.
For atoms with $I<E_{\rm Ps}=$6.8~eV, the $s-$type bound state contact densities are expected to be proportional to $\kappa=\sqrt{2 \varepsilon_b}$ with the proportionality constant of 0.66 determined previously \cite{RevModPhys.82.2557}.  
In this work, the proportionality constants are found to be about 0.35 for Be, 0.3 for Mg, and 0.4 for Zn, indicating that the present contact density results could be too small with respect to the corresponding binding energies.
This could be due to underestimated enhancement factors or slower convergence of contact densities with respect to the basis set size \cite{Bromley06,Ludlow10}.
We note that we are using $GW$ ionization energies in the enhancement factor formula instead of the HF ones, which are smaller in magnitude and would therefore lead to larger enhancement factors. 
The enhancement factors were also employed in recent model-potential calculations for Be and Mg obtaining good agreement with previous \emph{ab initio} calculations \cite{Swann:2020}.  

Concerning the normalization constant $a$,  its value is found to decrease with increasing binding energy.
The highest values of $a \approx 0.8$ are obtained for Be and Zn,  while Mg, with $\varepsilon_b$ about twice as large, has $a \approx 0.7$.
For Ca,  the value of $a$ is less than 0.5,  indicating that the most loosely bound electron is attached to the positron forming a Ps cluster, 
as expected based on the value of the ionization energy being lower than the Ps formation threshold.
Interestingly, the values of $a$ are usually lower for $p$-type bound states in Ca compared to $s-$ states,  but they are larger for Mg $p$-states at some levels of theory.

Finally, it is instructive to consider the dimensionless `strength parameter' $\mathcal{S} = - \sum_{\nu>0} \varepsilon_{\nu}^{-1}\langle\nu|\Sigma|\nu\rangle $ \cite{Dzuba:1994} (where the sum is over excited HF positron states of energy $\varepsilon_{\nu}$) that gives an effective measure of the strength of the self-energy $\Sigma$ evaluated at $E=\varepsilon_b$. 
The contribution of each occupied orbital to $\mathcal{S}$ can be calculated separately at the $\mathcal{S}^{(2)}$ and $\mathcal{S}^{\Gamma}$ levels (whose diagrams involve and can thus be labelled by a single hole).
The ratio of $\mathcal{S}^{\tilde{\Gamma}}$ and $\mathcal{S}^{(2)}$ varies from 0.8 (Zn) to 1.2 (Ca), as can be seen in Table~\ref{table:strpar}.
Notably, the ratio is considerably lower for polar molecules (about 0.5) \cite{Jaro22} suggesting a stronger relative contribution of the $\Gamma$ diagram for atoms compared to molecules.
The present method can quantify the individual orbital contributions to the positron self-energy (correlation potential) and all orbitals are treated on the same footing.
Evaluating $\mathcal{S}^{\tilde{\Gamma}}$ and $\mathcal{S}^{(2)}$ for each orbital,
the valence $s$-orbital is found to contribute more than 99\% of $\mathcal{S}^{\tilde{\Gamma}}$ in all cases except for Zn, for which the valence contribution is approximately 95\%.  
In the case of $\mathcal{S}^{(2)}$, the valence $s$-orbital is responsible for about 99,  97,  and 92\% of the correlation potential for Be, Mg, and Ca,  respectively, while for Zn it is only 83\%. 

\begin{table}[t!]
\centering
\caption{
Dimensionless strength parameter of the correlation potential for neutral atoms in different approximations to the positron self energy.
Positive (negative) strength parameters denote attractive (repulsive) positron-atom interactions. 
\label{table:strpar}}
\centering
\begin{ruledtabular}
\begin{tabular}{l c c c c c}
 &	 $\mathcal{S}$\textsuperscript{(2)}  &	 $\mathcal{S}$\textsuperscript{\textit{GW}}& $\mathcal{S}^{\tilde{\Gamma}}$ & $\mathcal{S}^{\tilde{\Lambda}}$ &$\mathcal{S}^{\textit{GW}+\tilde{\Gamma}+\tilde{\Lambda}}$ \\
 \hline\\[-1ex]
Be 	& 5.3 & 5.0 & 4.8& $-$0.9 & 8.9 	\\
Mg 	& 7.3 & 6.6 & 7.9& $-$1.3 & 13.2	\\
Zn 	& 6.8 & 6.0 & 5.4& $-$1.1 &	 10.3	\\
Ca 	& 10.2 & 9.2 & 12.5& $-$1.8 &	19.9	\\
\end{tabular}
\end{ruledtabular}
\end{table}%

\section{Summary and conclusion}
Many-body theory calculations of positron binding in several negative ions and neutral atoms were performed using a 
Gaussian-basis approach implemented in the EXCITON+ program \cite{Jaro22}. 
Good agreement with previous B-spline atomic many-body theory results of \citet{Ludlow10} was found for negative ions.
It was shown that for the ions, which bind the positron strongly at the mean-field HF level, the $GW$ and virtual-Ps diagrams alone produce binding energies within about 5\% of the most sophisticated approximation of the present method. 
In absolute terms, the inclusion of the positron-hole ladder series and the ladder series screening effects lower the binding energies by about 0.3--0.5 eV, becoming important for positron binding to the neutral atoms, where binding is generally weaker than in negative ions.
Notably, our results for Be, Mg, Zn, and Ca indicate that screening of the Coulomb interactions within virtual-Ps and positron-hole ladder series diagrams decreases binding energies more than the inclusion of (bare) positron-hole ladder series.
Moreover, individual orbital contributions to the positron self-energy were evaluated, providing fundamental insight that can instruct other methods.
Including lower valence orbitals in the calculations was shown to be important especially for larger atoms such as Zn. 
 
Our binding energy results are consistently about 0.2~eV higher than those obtained in the past using other theoretical methods.
From the available past results, reasonable agreement is found with coupled-cluster calculations of Harabati and coworkers \cite{Harabati14}, who suggest that their method likely overestimates true binding energies as a result of comparison with the stochastic variational calculation for Be, but offer no physical explanation, basing this only on the discrepancy with the result of the SVM calculation for Be \cite{Mitroy10}.
Overall, the lack of consensus in calculated binding energies warrants further theoretical attention. 
The present approach could be developed to include coupling of the three self-energy channels by consideration of the three-particle propagator via the Fadeev method \cite{Fadeevmol}, and a self-consistent calculation of electron and positron Green's functions. These are challenging computational problems, but could likely shed light on the discrepancies between approaches.  Comparison with measured scattering cross sections with calculations of positron binding energies and scattering cross sections using the same many-body framework (self-energy) could also shed light on the discrepancies. 
Ultimately, we hope this work will spur further theoretical and experimental studies on positron-atom bound states. 

\section*{Acknowledgements}
We thank David Waide, Andrew Swann, Jack Cassidy, Sarah Gregg and Gleb Gribakin for useful discussions, and Ian Stewart, Luis Fernandez Menchero (QUB), Martin Plummer and Alin Elena (UK STFC Scientific Computing Department) for high-performance computing assistance. This work was funded by the European Research Council grant 804383 ‘ANTI-ATOM’.


%

\end{document}